\documentclass[jcp,twocolumn]{revtex4-1}
\usepackage{amsmath}
\usepackage[english]{babel}
\usepackage{latexsym}
\usepackage{mathrsfs}
\usepackage{graphicx}
\usepackage{graphics}
\usepackage{color}
\usepackage{bm}
\usepackage{units,nicefrac}

\begin{document}

\title{Unified scaling law for rate factor of crystallization kinetics}

\author{Anatolii~V.~Mokshin} \email{anatolii.mokshin@mail.ru}
\affiliation{Kazan Federal University, Kazan, 420008 Russia}
\affiliation{Udmurt Federal Research Center of the Ural Branch of the Russian Academy of Sciences, 426067 Izhevsk, Russia}

\author{Bulat N. Galimzyanov} \email{bulatgnmail@gmail.com}
\affiliation{Kazan Federal University, Kazan, 420008 Russia}
\affiliation{Udmurt Federal Research Center of the Ural Branch of the Russian Academy of Sciences, 426067 Izhevsk, Russia}

\author{Dinar T.~Yarullin}
\affiliation{Kazan Federal University, Kazan, 420008 Russia}

\begin{abstract} 
	Features of the crystallization kinetics define directly the rate characteristics: the crystal nucleation rate, the crystal growth rate and the so-called kinetic rate factor known also as the attachment rate (of particles to the surface of a crystalline nucleus). We show that the kinetic rate factor as function of the reduced temperature follows a unified scaled power law. This scenario is confirmed by our simulation results for model atomistic systems (crystallizing volumetric liquids and liquid thin film) and by available experimental data for crystallizing polymers. We find that the exponent of this unified scaling law is associated with a measure of the glass-forming ability of a system. The results of the present study extend the idea of a unified description of the rate characteristics of the crystal nucleation and growth kinetics by means of the scaling relations.
\end{abstract}

\maketitle

\section{Introduction}\label{intro}

Crystallization is a typical first-order phase transition, the time scale of which is determined by such the rate characteristics as the nucleation rate $J_s$, the growth rate $v_s$ and the kinetic rate factor $g^+$ referred also to as the attachment rate~\cite{Kashchiev_2000,Schmelzer_2015,Kelton_Greer_2010,Kalikmanov_2012,Turci_2014,Song_Mendelev_2018}. Among these rate characteristics, the kinetic rate factor is of special interest for a number of reasons. First of all, this quantity is the main input parameter for many theories of nucleation and growth, including the Becker-D\"oring gain-loss theory, within the framework of which a theoretical description of nucleation and growth processes was first implemented~\cite{Weinberg_2002,Mokshin_2017,Turnbull_Fisher_1949,Kelton_Greer_1986}. Secondly, the kinetic rate factor $g^+$ accounts for the attachment of particles to a nucleus of an emerging (crystalline) phase~\cite{Kashchiev_2000}. Therefore, evaluation of $g^+$ can be necessary to determine the nucleus shape for the peculiar case of anisotropic nucleus growth~\cite{Galenko_2019}. As an example, one can mention the studies of the ice crystal growth given in Refs.~\cite{Demange_2017_1,Demange_2017_2,Barrett_2012}. According to the basic definition, the attachment rate $g^+$ is scalar quantity and it does not account for the geometry of the growing surface. This quantity is directly related with the `diffusion’ of the nuclei along the size axis (see pages 126-127 in Ref.~\cite{Kashchiev_2000}). It is clear that values of $g^+$ are dependent on type of the considered system, on the thermodynamic state (e.g., supercooling level), on the nucleus size and the crystal growth mode. However, by analogy with the self-diffusion coefficient, for the crystal growth at any  thermodynamic state, one can define just the most probable value of $g^+$. This point is discussed also in detail in Ref.~\cite{Auer_Frenkel_2004}. Finally, there are still no experimental methods for \textit{direct measurements} of the rate factor $g^+$. One of the used ways to evaluate this term empirically is to identify the quantity $g^+$ with the experimentally measured diffusion coefficient, the viscosity coefficient and other relevant kinetic parameters.
\begin{figure*}[ht!]
	\centering\includegraphics[width=0.8\linewidth]{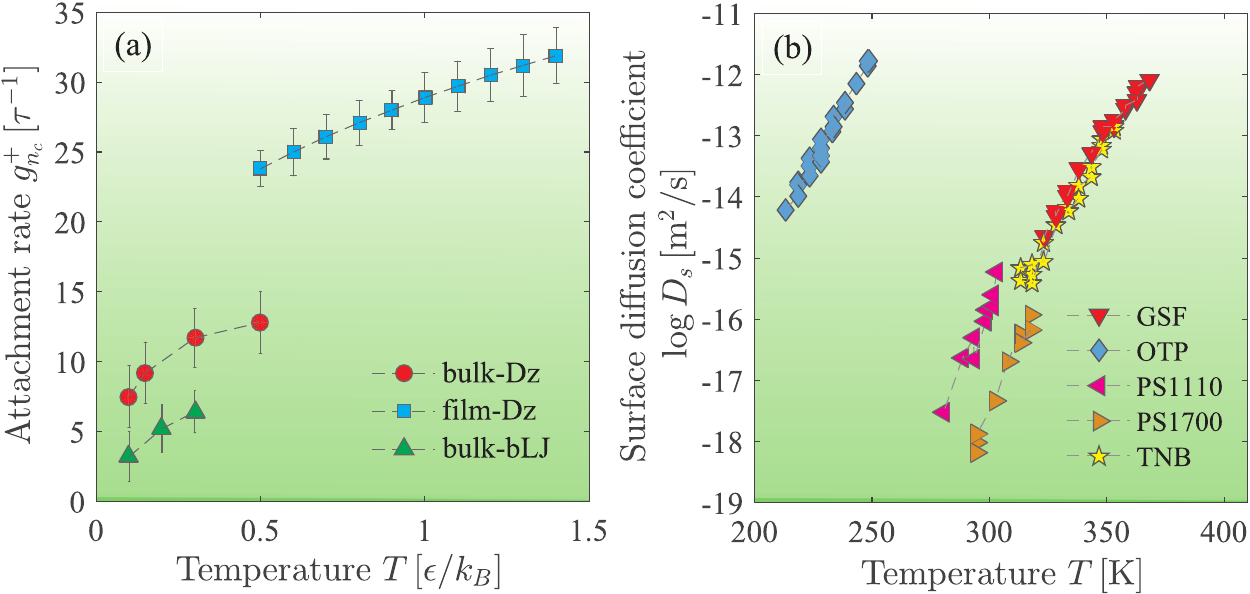}
	\caption{(color online) (a) Rate factor $g_{n_{c}}^{+}$ as function of the temperature $T$ for the simulated systems: the bulk-Dz~\cite{Mokshin_2017}, the bulk-bLJ~\cite{Mokshin_2015}, the film-Dz~\cite{Galimzyanov_Mokshin_2019}. Here, the quantities $g_{n_c}^{+}$ and $T$ are given in units of $\tau^{-1}$ and $\epsilon/k_{B}$, where $k_{B}$ is the Boltzmann constant. (b) Experimental data for the surface diffusion coefficient $D_{s}$ for different crystallizing polymers~\cite{Huang_2017}.}
	\label{fig:1}
\end{figure*}
It is expected that reliable temperature dependence of the attachment rate $g^+(T)$ can be obtained from the experimental data for the surface diffusion $D_s(T)$ within the following approximation:
\begin{equation}\label{Approx}
g^{+}(T) = C \ D_s(T).
\end{equation}
where the coefficient $C$ has a dimension of (length)$^{-2}$ and is associated with the diffusion length. Relation~(\ref{Approx}) is capable to provide only a qualitative estimate of the attachment rate $g^+(T)$. Moreover, relation~(\ref{Approx}) is valid when the particle diffusion is not driven by external fields \cite{Schmelz_2003} and when the most probable attachment rate for the thermodynamic state is considered.

It this work, we compute the rate factor $g^+(T)$ for the three model atomistic crystallizing systems -- the volumetric binary Lennard-Jones liquid (bulk-bLJ), the volumetric Dzugutov liquid (bulk-Dz) and the model liquid thin film (film-Dz). Note that the quantity $g^+(T)$ for the considered systems is determined directly from the nucleus growth trajectories computed on the basis of the molecular dynamics simulations results; and, therefore, no any approximations were applied to determine $g^+(T)$. We compare our results with the experimental data and exam the idea of unified scaling laws for the rate characteristics of the crystallization kinetics~\cite{Huang_2017}.

\section{Kinetic rate factor \textit{vs.} temperature} \label{sec:1}

Let us consider a liquid which is supercooled to a some thermodynamic state with a temperature $T$; and $T<T_m$, where $T_m$ is the melting temperature. For a crystalline nucleus emerging and growing in this system, the kinetic rate factor $g^+$ will depend on the nucleus size $n$~\cite{Song_Mendelev_2018}. We shall restrict our consideration of the quantity $g^+$ to the case of the nucleus of the critical size $n_c$ and we shall evaluate the  rate factor $g_{n_{c}}^{+}$ of the particle attachment to the surface of the $n_c$-sized nucleus~\cite{Galimzyanov_Mokshin_2019}.

Further, let us assume that the set of the growth trajectories $n(t)$ of a growing nucleus are determined for the time window $t\in[\tau_{c}-\tau_{w}; \tau_{c}+\tau_{w}]$, which defines the vicinity of the waiting time $\tau_c$ for a critically-sized nucleus; $\tau_{w}$ is the half-width of the time window. The growth trajectories can be computed, for example, from molecular dynamics simulations results for the system. Then, the rate factor $g_{n_{c}}^{+}$ can be determined on the basis of the known set of the growth trajectories $n(t)$ as follows~\cite{Auer_Frenkel_2004}:
\begin{equation}\label{eq_3}
g_{n_{c}}^{+}=\frac{1}{2} \frac{\left\langle\left[n(t)-n_{c}\right]^{2}\right\rangle}{t}\Bigg|_{t\in[\tau_{c}-\tau_{w}; \tau_{c}+\tau_{w}]}.
\end{equation}
Here, the angle brackets $\left\langle...\right\rangle$ denote an averaging over set of the growth trajectories.
We compute $g_{n_{c}}^{+}$  for our systems on the basis of the molecular dynamics simulation results and we take the parameter $\tau_{w}=10\,\tau$ to use Eq.~(\ref{eq_3}); here, $\tau=\sigma \sqrt{m/\epsilon}$ is the time unit, $m$ is a particle mass, $\sigma$ is a particle diameter and $\epsilon$ is the unit energy~\cite{Mokshin_2017}. Details of the molecular dynamics simulations are given in Refs.~\cite{Mokshin_2017,Galimzyanov_Mokshin_2019}.

Figure~\ref{fig:1}a shows the quantity $g_{n_{c}}^{+}$ as function of the temperature $T$ computed for the crystallizing model liquids: the bulk-bLJ, the bulk-Dz and the film-Dz.
As seen, the function $g_{n_{c}}^{+}(T)$ increases with the temperature for all the simulated systems. This scenario agrees qualitatively  with available experimental data for the surface diffusion coefficient $D_{s}(T)$~\cite{Huang_2017} evaluated for crystallizing griseofulvin (GSF), ortho-terphenyl (OTP),  polystyrene oligomers (PS1110) and (PS1700), tris-naphthyl benzene (TNB) (see Fig.~\ref{fig:1}b). Both the quantities $g_{n_{c}}^{+}(T)$ and $D_{s}(T)$ are measured in various physical units. Namely, the rate $g_{n_{c}}^{+}(T)$ is measured in units of (time)$^{-1}$, whereas the coefficient $D_{s}(T)$ has a dimension of (length)$^2$/(time).  Nevertheless, taking into account relation~(\ref{Approx}) and results of Fig.~\ref{fig:1}, one can reasonably assume that the rate factor as well as the surface diffusion coefficient can obey a common unified scaling law.
\begin{figure}[h!]
	\centering\includegraphics[width=0.8\linewidth]{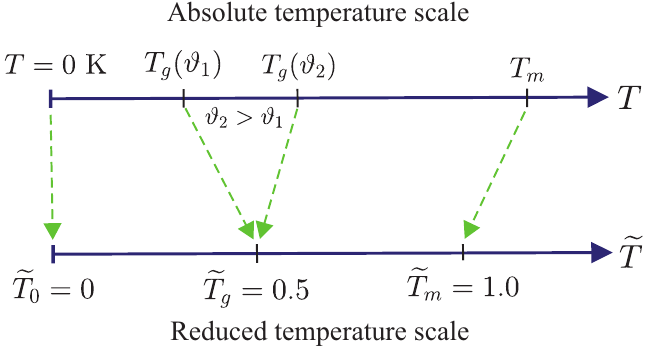}
	\caption{(color online) Schematic plot of the correspondence between the absolute temperature scale $T$  and the reduced temperature scale $\widetilde{T}$. The quantities $T_0$, $T_g$ and $T_m$ are the zeroth temperature, the glass transition temperature and the melting temperature, respectively.  Note that the glass transition temperature $T_g$ depends on the cooling rate $\vartheta$  for the absolute temperature scale, and $T_g(\vartheta_{2})>T_g(\vartheta_{2})$ at $\vartheta_{2}>\vartheta_{1}$. The glass transition temperature takes the fixed value $\widetilde{T}_g=0.5$ for the $\widetilde{T}$-scale.}
	\label{fig:2}
\end{figure}
Crystallization of supercooled liquids and glasses proceeds at the temperatures from the range $0 < T\leq T_m$. For an isobar, this temperature range contains three critical temperatures: the zeroth temperature $T_{0}=0\,$K; the glass transition temperature $T_g$, and the melting temperature $T_m$. As discussed in detail before in Refs.~\cite{Mokshin_2015,Avramov_Zanotto_2003,Nascimento_2007}, it is not possible to take into account unified regularities of the crystallization characteristics as dependent on the temperature, if we use the absolute temperature scale $T$ or the reduced temperature scales $T/T_g$ and $T/T_m$.

On the other hand, it was introduced in Ref.~\cite{Mokshin_2015} the reduced temperature scale $\widetilde{T}$, according to which the zeroth temperature $T_{0}$, the glass transition temperature $T_{g}$ and the melting temperature $T_{m}$ will take the fixed values for any system~\cite{Mokshin_2017,Mokshin_2015}: $\widetilde{T}_{0}=0$; $\widetilde{T}_g=0.5$; $\widetilde{T}_m=1$. If values of the temperatures $T_{m}$ and $T_{g}$ are known for a concrete system, then the reduced temperature scale $\widetilde{T}$ for this system is defined by relation (see also Fig.~\ref{fig:2}):
\begin{equation}\label{eq_1}
\widetilde{T}=\left[\frac{0.5-\left(\displaystyle{\frac{T_g}{T_m}}\right)^2}{1-\displaystyle{\frac{T_g}{T_m}}}\right]\left(\displaystyle{\frac{T}{T_g}}\right)+\left[\frac{\displaystyle{\frac{T_g}{T_m}}-0.5}{\displaystyle{\frac{T_m}{T_g}}-1}\right]\left(\frac{T}{T_g}\right)^2.
\end{equation}

Following Refs.~\cite{Mokshin_2017,Mokshin_2015}, we now take scaling relation for the rate factor as the next function of the reduced temperature:
\begin{equation}\label{eq_pl1}
	\frac{g_{n_{c}}^{+}(\widetilde{T})}{g_{n_{c}}^{(g)}}=\left(\frac{\widetilde{T}}{\widetilde{T}_{g}}\right)^{\chi}.
\end{equation}
Here, $g_{n_c}^{(g)}$ is the rate factor at the glass transition temperature $T_{g}$; the exponent $\chi$ is the positive adjustable parameter, which can be associated with a measure of the glass forming ability of a system (see discussion in Ref.~\cite{Mokshin_2015}). The smaller value of the parameter $\chi$, for a longer time a system is capable to keep a glassy state. It is necessary to note that if approximation~(\ref{Approx}) is fulfilled, then we have \[
\frac{g_{n_{c}}^{+}(\widetilde{T})}{g_{n_{c}}^{(g)}} = \frac{D_s(\widetilde{T})}{D_s^{(g)}},
\]
and, therefore, the same relation (\ref{eq_pl1}) holds for the surface diffusion coefficient $D_s(\widetilde{T})$ scaled to its value $D_{s}^{(g)}$ at the glass transition temperature. To compare data for the rate factor for the various systems, it is convenient to present these data in double logarithmic scale, for which relation~(\ref{eq_pl1}) should take the next form:
\begin{equation}\label{eq_6}
\frac{1}{\chi}\log\left [\frac{g_{n_c}^{+}(\widetilde{T})}{g_{n_c}^{(g)}}\right ]=\log\left [\frac{\widetilde{T}}{\widetilde{T}_g}\right].
\end{equation}

\begin{figure}[h!]
	\centering\includegraphics[width=1.0\linewidth]{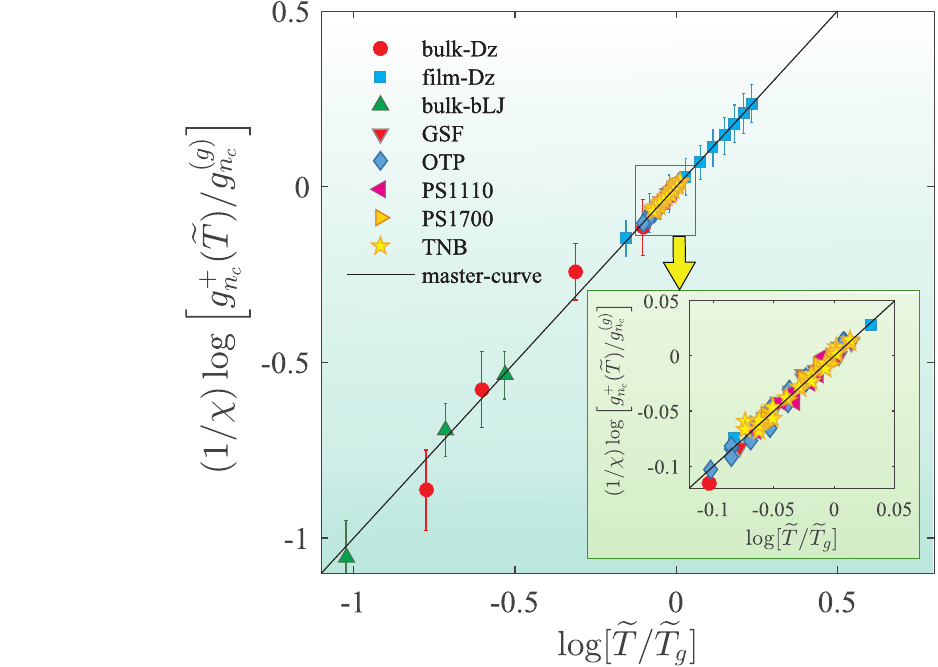}
	\caption{Scaled plot of the data for the rate factor for the various systems (the same with Fig.~\ref{fig:1}). The solid line indicates simple linear dependence resulted from Eq.~(\ref{eq_6}) with $\chi=1$.}
	\label{fig:3}
\end{figure}

Simple linear dependence will be in the scaling plot corresponding to relation~(\ref{eq_6}), whereas the slope of this linear dependence is regulated by the exponent $\chi$.
\begin{table}[h]
	\caption{Parameters of the considered systems required to perform the scaling~(\ref{eq_6}): glass transition temperature $T_{g}$; melting temperature $T_{m}$; exponent $\chi$; rate factor $g_{n_c}^{(g)}$ and index of fragility $m$.}
	\label{tab:1}
	\begin{tabular}{llllll}
		\hline\noalign{\smallskip}
		System & $T_g$ & $T_m$ & $\chi$ & $g_{n_c}^{(g)}$ & $m$ \\
		\noalign{\smallskip}\hline\noalign{\smallskip}
		bulk-Dz & $0.65\,\epsilon/k_B$ 	& $1.51\,\epsilon/k_B$ 	& $0.58\pm0.06$ & $15.2\,\tau^{-1}$ & --  \\
		film-Dz & $0.78\,\epsilon/k_B$ 	& $1.72\,\epsilon/k_B$ 	& $0.34\pm0.03$ & $24.7\,\tau^{-1}$ & --  \\
		bulk-bLJ & $0.92\,\epsilon/k_B$ 	& $1.65\,\epsilon/k_B$ 	& $0.31\pm0.04$ & $16.5\,\tau^{-1}$ & --  \\
		GSF 			& $361$ K				& $493$ K				& $21\pm3$ 		& -- & $84.6$~\cite{Shi_Zhang_2017} \\
		OTP 			& $246$ K 				& $331$ K				& $14\pm2$ 		& -- & $81$~\cite{Miriglan_Schweizer_2014} \\
		PS1110			& $307$ K 				& $513$ K				& $45\pm4$ 		& -- & $\simeq140$~\cite{Santangelo_Roland_1988} \\
		PS1700 			& $320$ K 				& $533$ K				& $45\pm4$ 		& -- & $\simeq140$~\cite{Santangelo_Roland_1988} \\
		TNB  			& $347$ K 				& $467$ K				& $20\pm3$ 		& -- & $84$~\cite{Miriglan_Schweizer_2014} \\
		\noalign{\smallskip}\hline
	\end{tabular}
\end{table}

In Fig.~\ref{fig:3}, we show the data for the rate factor scaled according to Eq.~(\ref{eq_6}). As seen, all the data collapse into the unified linear dependence. We find that for simulated atomistic systems the exponent is $\chi<1$, whereas for the considered molecular glasses the exponent takes values within the range from $\chi\simeq14$ (for OTP) to $\chi\simeq45$ (for PS oligomers) [see Table~\ref{tab:1}]. Remarkably, value of the exponent $\chi$ depends on type of the system and there is a correlation between the exponent and the Angell's fragility index $m$~\cite{Angell_Ngai_2000}. The larger value of the exponent $\chi$, the larger value of the index $m$ (see Table~\ref{tab:1}). For example, for OTP, the exponent $\chi\simeq14$ corresponds to the fragility index $m=81$~\cite{Miriglan_Schweizer_2014}, whereas the exponent $\chi$ takes value $45$ for the more fragile polystyrene oligomers (PS1110) and (PS1700) whose the index fragility $m$ is $\simeq140$~\cite{Santangelo_Roland_1988}.

\section{Concluding remarks}\label{concl}

There are features of crystallization that are common to all systems, regardless of their specific type. Namely, with increase of supercooling level, the viscosity and the chemical potential difference increase, whereas the particle mobility decreases. Further, application of the classical nucleation theory is not restricted to specific systems; the basic equation of the KJMA-theory for crystallization has universal character. So, some features of the crystallization kinetics will be manifested in a common manner for all the systems. The kinetic rate factor of crystal nucleation and crystal growth is associated directly with the particle mobility, and expectations about some unified scenario with this quantity for all the systems should be quite reasonable. The difference for systems of different types (metals, polymers, network-systems etc) will be manifested in the magnitude of the change for the considered quantity (the kinetic rate factor) on the same scaled temperature range.

In this work, on the basis of the molecular dynamics simulations we evaluated the temperature dependence of the kinetic rate factor for various model crystallizing liquids and compared these results with the experimental data. We found that the rate factor $g_{n_c}^{+}$ and the surface diffusion $D_s$ as functions of the reduced temperature $\widetilde{T}$ follow a unified power-law dependence, where the exponent $\chi$ can be associated with the measure of the glass-forming ability of systems. These results support the idea of unified scaling laws for the rate characteristics of the crystallization kinetics.

\begin{acknowledgments}
\noindent This work is supported by the Russian Science Foundation (project No.~19-12-00022).
\end{acknowledgments}

\end{document}